\documentclass{vgtc}

\graphicspath{{figures/}{pictures/}{images/}{./}}

\usepackage{times}

\usepackage{booktabs}
\usepackage{mathptmx}
\usepackage{xspace}
\usepackage{amsmath}
\usepackage{amssymb}
\usepackage{xcolor}

\onlineid{0}

\vgtccategory{Research}

\vgtcinsertpkg

\usepackage{scalerel}
\usepackage{tikz}
\usetikzlibrary{svg.path}
\definecolor{orcidlogocol}{HTML}{A6CE39}
\tikzset{
  orcidlogo/.pic={
    \fill[orcidlogocol] svg{M256,128c0,70.7-57.3,128-128,128C57.3,256,0,198.7,0,128C0,57.3,57.3,0,128,0C198.7,0,256,57.3,256,128z};
    \fill[white] svg{M86.3,186.2H70.9V79.1h15.4v48.4V186.2z}
      svg{M108.9,79.1h41.6c39.6,0,57,28.3,57,53.6c0,27.5-21.5,53.6-56.8,53.6h-41.8V79.1z M124.3,172.4h24.5c34.9,0,42.9-26.5,42.9-39.7c0-21.5-13.7-39.7-43.7-39.7h-23.7V172.4z}
      svg{M88.7,56.8c0,5.5-4.5,10.1-10.1,10.1c-5.6,0-10.1-4.6-10.1-10.1c0-5.6,4.5-10.1,10.1-10.1C84.2,46.7,88.7,51.3,88.7,56.8z};
  }
}
\newcommand{\authororcid}[2]{%
  \texorpdfstring{%
    \href{https://orcid.org/#2}{#1 \mbox{\scalerel*{%
      \begin{tikzpicture}[yscale=-1,transform shape]%
        \pic{orcidlogo};%
      \end{tikzpicture}%
    }{|}}}%
  }{#1}%
}

\newcommand{\dtour}{\texorpdfstring{\texttt{dtour}\xspace}{dtour}}

\title{dtour: A Steerable \emph{Tour de Vis} Through High-Dimensional Data}

\teaser{
  \centering
  \includegraphics[width=\linewidth]{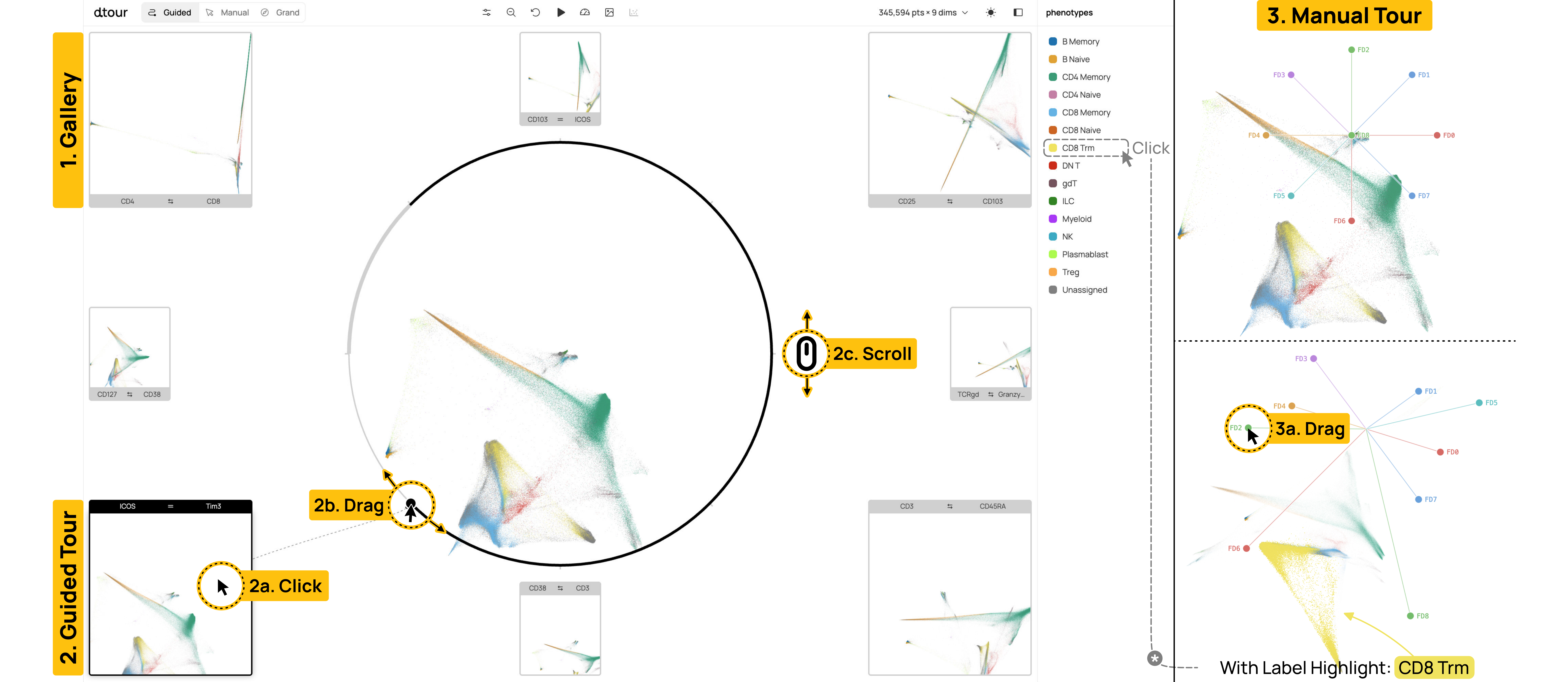}
  \caption{
    \dtour{}'s interface for exploring high-dimensional data as a tour of keyframe projections, unifying three modes of increasing traversal steerability. (1) A central 2D scatter ringed by projection previews gives an overview. (2) Clicking a preview, scrubbing the slider, or scrolling advances the scatter along a cyclical guided tour, smoothly transitioning between projections. (3) For details, the user can switch to manual axis manipulation and highlight points by label or lasso selection.
  }
  \label{fig:teaser}
}

\author{\authororcid{Fritz Lekschas}{0000-0001-8432-4835}\thanks{e-mail: fritz@ridgedata.ai}\\ %
        \scriptsize Ridge AI %
\and \authororcid{Nezar Abdennur}{0000-0001-5814-0864}\thanks{e-mail: nezar.abdennur@umassmed.edu}\\ %
     \scriptsize UMass Chan Medical School %
}

\abstract{
Understanding high-dimensional data requires projecting it into lower-dimensional spaces, but any single projection inevitably loses information or introduces distortions. Tours address this limitation through animation of 2D projection sequences, yet existing tools present tradeoffs in the freedom and steerability of projection traversal, providing little to no ability to move between expert-guided paths and unrestrained exploration. We present \dtour{}, a tour interface that combines static projection previews, reversible scrubbing along continuous geodesic projection paths, manual projection manipulation, and a wandering grand tour, all within a single progressive exploration interface. \dtour{} scales to millions of points via GPU-accelerated rendering, runs in any modern browser, and integrates with both Python and JavaScript ecosystems. We demonstrate \dtour{} on text, image, and single-cell data for two usage scenarios: gradually revealing structure in high-dimensional data and validating non-linear dimensionality reduction outputs.
}


\begin{document}

\firstsection{Introduction}
\label{sec:intro}

\maketitle

Understanding high-dimensional data is fundamentally challenging as human perception is limited to three dimensions and visual exploration necessarily involves projecting data into lower-dimensional spaces. Linear dimensionality reduction (DR) methods like Principal Component Analysis (PCA) preserve global structure faithfully, but any single projection hides all structure orthogonal to the chosen projection plane. By contrast, non-linear neighbor-based approaches like t-Distributed Stochastic Neighbor Embedding (t-SNE~\cite{van2008visualizing}) and Uniform Manifold Approximation and Projection (UMAP~\cite{mcinnes2018umap}) attempt to capture manifold structure in a single lower-dimensional space, but inevitably introduce distortions that can misrepresent cluster structure and neighborhood relationships. Despite many debates about the usefulness and faithfulness of non-linear DR methods (e.g.~\cite{chari2023specious,lause2024art}), such methods are widely used and undoubtedly useful if interpreted with care~\cite{wattenberg2016how,kobak2019art,becht2019dimensionality}.

Presenting more than a single 2D or 3D projection helps interpret both linear and non-linear DR outputs. Approaches such as scatter plot matrices~\cite{chambers1983graphical} and small multiples~\cite{tufte1983visual} lay out a small set of fixed projections side by side. Alternatively, \emph{tours}~\cite{asimov1985grand,buja2005computational} present projections sequentially as an animated path through projection space~\cite{swayne2001ggobi,wickham2011tourr}. Tour variants differ in how the path is generated: \emph{grand} tours traverse the space by a random walk~\cite{asimov1985grand,buja1986grand}, \emph{guided} tours~\cite{cook1995grand} select specific target projections by optimizing an interestingness criterion such as cluster separation, and \emph{manual} tours~\cite{cook1997manual} give control of the projection to the user~\cite{li2020visualizing}.%
\footnote{Throughout, we use ``guided tour'' to refer to any precomputed sequence of keyframe projections that the user traverses along a fixed path, regardless of how the keyframes were selected.
}

Approaches for touring multiple projections lie along a spectrum of \emph{freedom of traversal} and differ in user \emph{steerability}. At one end, grids of static projections are directly comparable at a glance but require constant shifts of focus to integrate information and scale poorly with the number of projections shown. In the middle, animated tours~\cite{swayne2001ggobi,wickham2011tourr} eliminate focus shifts and help keep track of correspondences by morphing between projections in a single view. Playback lets the user pace the traversal along a fixed precomputed path, but only a single projection is visible at a time. At the other end, manual tours give full control, but reaching an informative projection by hand is slow and cognitively demanding. These trade-offs are unavoidable within any single tour mode, but they can be reconciled by an interface that lets the user move smoothly across the spectrum itself.

Here we present \dtour{}, a tour interface for high-dimensional data designed to provide frictionless control over the freedom and steerability of projection traversal. Given data with four or more dimensions, \dtour{} opens with a central 2D scatter surrounded by a gallery of \emph{keyframe} projection previews (\autoref{fig:teaser}.1). The user advances the central scatter between keyframes by clicking previews or smoothly transitions along the path connecting them as a guided cyclical tour via scrubbing or scrolling (\autoref{fig:teaser}.2). When more precise control is needed, manual manipulation (\autoref{fig:teaser}.3) lets the user directly control the influence of individual axes on the central projection, allowing user-driven excursions from the primary tour. Complementing these modes, a grand tour animates a random walk through projection space for hands-off serendipitous exploration. All transitions---within a tour mode and between---are smoothly interpolated to preserve point identity and spatial context. 
Combined with lasso selection and label-based highlighting (\autoref{fig:teaser} *), \dtour{} provides the fluid, progressive control over traversal complexity needed to navigate high-dimensional data effectively.

We implemented \dtour{} as a general-purpose tool that scales to data with millions of points through GPU-accelerated rendering. \dtour{} is agnostic to how keyframe projections are produced: they can come from a projection pursuit, hyperparameter sweeps, a time series, or any other source that yields a sequence of 2D projections with one-to-one point correspondence. We demonstrate two usage scenarios: (1)~revealing structure in high-dimensional data and (2)~validating non-linear dimensionality reduction outputs with text, image, and single-cell datasets.  \dtour{} runs in any modern browser (\url{https://dtour.dev}), is available as a widget for Jupyter and Marimo notebooks, and can be embedded in React applications or built upon via its rendering engine. The source code is available at \url{https://github.com/flekschas/dtour}.

\section{Related Work}
\label{sec:related-work}

\subsection{Dimensionality Reduction}
\label{sec:dimensionality-reduction}

Linear dimensionality reduction methods such as PCA project data onto variance-preserving subspaces, yielding interpretable axes but capturing only linear structure. Non-linear techniques---notably \mbox{t-SNE}~\cite{van2008visualizing} and UMAP~\cite{mcinnes2018umap}---aim to capture manifold structure in a single layout, but inevitably introduce distortions~\cite{chari2023specious}. B{\"o}hm et al.~\cite{boehm2022attraction} show that these and other neighbor-embedding methods lie on a spectrum of attraction between neighbors and repulsion between all points, where this balance trades off continuous structure against cluster separation. The MDE framework~\cite{agrawal2021mde} further unifies linear and non-linear objectives into a single optimization formulation, enabling systematic comparison across the DR spectrum. Whether linear or non-linear, a fixed 2D view cannot fully represent high-dimensional data, motivating the use of tours to examine data from multiple perspectives.

\subsection{Visualization of Embeddings}
\label{sec:embedding-visualization}

A growing ecosystem supports interactive exploration of embedding projections. The Embedding Projector~\cite{embedding_projector} was an early web-based system for browsing embeddings with nearest-neighbor search and support for PCA, t-SNE, and custom projections. Scalability has since been a central concern: tools such as regl-scatterplot~\cite{lekschas2023regl}, Jupyter Scatter~\cite{lekschas2024jupyter}, WizMap~\cite{wang2023wizmap}, DataMapPlot~\cite{mcinnes2024datamapplot}, and Embedding Atlas~\cite{ren2025embedding_atlas} can render millions of points with interactive visual encodings, selections, and pan-and-zoom navigation. Emblaze~\cite{sivaraman2022emblaze} and Comparative Embedding Visualization~\cite{manz2024general} support comparison across multiple embedding spaces, but are limited to pairwise views. All of these systems present one or two static 2D projections; none offer smooth, steerable traversal through projection space or across embedding sequences.

\subsection{Tour Methods}
\label{sec:tour-methods}

Animating sequences of low-dimensional projections dates back to Asimov's grand tour~\cite{asimov1985grand}, a smooth random walk through all possible 2D projections. The mathematical foundations for these dynamic projections, including geodesic interpolation on the Stiefel manifold and the general framework of $d$-dimensional projections from $p$-dimensional space, were formalized by Buja et al.~\cite{buja2005computational}. Cook et al.~\cite{cook1995grand} introduced \emph{guided tours}, which replace random target selection with projection pursuit optimization, steering the animation toward projections that maximize a criterion of interestingness (e.g., holes, central mass, or linear discriminant indices). Additional variants include local tours, which rock near a given projection, and the manual tour~\cite{cook1997manual} for direct variable control. A comprehensive review of tour methods is given by Lee et al.~\cite{lee2022tours}.

\subsection{Tour Software and Applications}
\label{sec:tour-software}

The primary tour ecosystem is the R package \texttt{tourr}~\cite{wickham2011tourr}, which implements grand, guided, local, manual, and other tour types with multiple displays. Earlier interactive systems include GGobi~\cite{swayne2003ggobi} and its predecessor XGobi, which pioneered linked brushing and direct manipulation of projections. Recent packages extend this ecosystem with refined manual controls~\cite{spyrison2020spinifex}, linked DR diagnostics~\cite{lee2020liminal}, portable HTML rendering~\cite{hart2022detourr}, and Langevin-dynamics-based smooth paths~\cite{harrison2023langevitour}. In an interactive article, Li et al.~\cite{li2020visualizing} present grand and manual tours with steerable axes within a single interface to visualize neural-network activations.
Yet these tools limit accessibility as the analyst must choose the most suitable mode upfront. \dtour{} addresses this gap with a single interface that unifies overview, guided, manual, and grand tour modes across the steerability spectrum, while scaling to million-point datasets.

\section{The \dtour{} Method}
\label{sec:dtour}

\dtour{} is designed to support two tasks that no single 2D projection can serve: \emph{revealing} structure in high-dimensional data by examining it across many projections, and \emph{validating} whether structure in an embedding is genuine or a projection artifact. Both require following how structure changes across projections, driving \dtour{}'s core design principle: smooth, user-steered traversal within and between projection spaces.

\subsection{User Interface}
\label{sec:user-interface}

Inspired by Shneiderman's visual information-seeking mantra~\cite{shneiderman2003eyes}, \dtour{} supports progressive exploration of high-dimensional data through a unified interface (\autoref{fig:teaser}): an overview consisting of a central 2D scatter surrounded by a gallery of keyframe projection previews, a guided tour that smoothly animates through the keyframe sequence, and a manual tour that lets the user manipulate the projection by dragging dimension axes. Complementing these modes, a grand tour mode enables random rotations through projection space as a continuous playback. The central scatter serves as a fixed reference point across all modes, and all projection changes are smoothly interpolated to preserve object constancy, letting viewers track points across views rather than re-identifying them after each transition~\cite{robertson1993information,rodrigues2024comparative}.

\paragraph{Keyframe gallery.}
On launch, a gallery of previews (\autoref{fig:teaser}.1) surrounds the central scatter, one per keyframe, each showing the data in that keyframe's projection. Clicking a preview (\autoref{fig:teaser}.2a) advances the central view to that keyframe. Tours can specify feature loadings, shown as text beneath each preview indicating the top contributing dimensions. Gallery previews are arranged around the circular tour slider (described below), such that the gallery acts as a lookahead and an orientation device.

\paragraph{Guided tour.}
A circular tour slider (\autoref{fig:teaser}.2b) controls the current position along the arc-length-parameterized keyframe path (\autoref{sec:interpolation}). Users can scrub the slider, scroll the mouse wheel, or press play for animated playback to advance the tour. Tick marks at keyframe positions serve as navigational landmarks, and the width of the slider's ring segments encode geodesic distances between consecutive keyframes: thin segments indicate stretched regions of projection space, thick segments indicate compressed regions. The tour runs as a closed loop so that continuous forward or backward traversal never encounters a discontinuity.

\paragraph{Manual tour.}
In manual mode, dimension axes appear as draggable handles (\autoref{fig:teaser}.3) overlaid on the scatter plot, one per data dimension. Each handle's projected direction and length encode that dimension's current contribution to the projection basis, doubling as a control surface and a feature-loading legend. Dragging a handle (\autoref{fig:teaser}.3a) specifies a new target direction for its variable and the remaining basis is re-orthonormalized to preserve a valid tour frame. Additionally, holding \texttt{Shift} while dragging rotates the view about a temporary third axis (the residual principal component) to help build spatial intuition. Together, these let the user isolate individual dimensions' effects.

\paragraph{Color encoding and point selection.}
Many datasets include labels or non-embedded dimensions that aid interpretation. These can be mapped to point color in \dtour{} via continuous, 2D, or categorical encodings. Lasso and label-based selection let users isolate a subset of points and track them across projections, enabling a \emph{select-then-explore} workflow (\autoref{fig:teaser} *): select points of interest during guided playback, then switch to manual mode to investigate which dimensions distinguish them.

\subsection{Tour Interpolation}
\label{sec:interpolation}

A tour is defined by a cyclic sequence of \emph{keyframe} projections, each represented as a $p \times 2$ orthonormal basis matrix $\mathbf{F}_i$ mapping $p$-dimensional data to 2D. Smoothly animating between keyframes requires a meaningful distance on the space of bases and an interpolation scheme that preserves orthonormality.

\paragraph{Geodesic distance.}
The distance between two 2D subspaces spanned by bases $\mathbf{F}_a$ and $\mathbf{F}_z$ is measured via the principal angles $\tau_0, \tau_1$ obtained from the singular value decomposition of $\mathbf{F}_a^{\!\top} \mathbf{F}_z$:
\begin{equation}
  d(\mathbf{F}_a, \mathbf{F}_z) = \sqrt{\tau_0^2 + \tau_1^2},
  \quad \tau_i = \arccos(\sigma_i)
  \label{eq:geodesic}
\end{equation}
where $\sigma_0, \sigma_1$ are the singular values clamped to $[-1,1]$. This is the geodesic distance on the Grassmannian of 2D subspaces~\cite{buja2005computational}. We use it only to pace playback, not to define the interpolation path. For the $2 \times 2$ case, \dtour{} computes the SVD analytically.

\paragraph{Catmull-Rom spline with re-orthonormalization.} \dtour{} interpolates the keyframe \emph{frames} directly. Given four consecutive bases $\mathbf{P}_0, \ldots, \mathbf{P}_3$, the interpolated basis at parameter $t \in [0,1]$ between $\mathbf{P}_1$ and $\mathbf{P}_2$ is the standard cubic Catmull-Rom~\cite{catmull1974class} blend applied element-wise, followed by Gram-Schmidt orthonormalization. Unlike a B\'ezier curve, which passes only through its end control points, Catmull-Rom interpolates every keyframe exactly, and its $C^1$-continuous tangents keep the animated path smooth across segments. We adopt it over spherical-linear or piecewise-geodesic interpolation, which are only $C^0$: their per-segment arcs meet at keyframes with mismatched tangents, producing a visibly abrupt change of direction each time the looping tour crosses a keyframe.

\paragraph{Arc-length parameterization.} To ensure perceptually uniform playback speed, \dtour{} precomputes a cumulative arc-length table by sampling each spline segment at eight interior points and summing geodesic distances between consecutive samples. At runtime, a binary search maps $t \in [0,1]$ to the correct segment and local parameter in $O(\log n)$ time, so that scrubbing the circular slider produces constant angular velocity through projection space.

\subsection{Tour Strategies}
\label{sec:tour-strategies}

\dtour{} accepts any sequence of $p \times 2$ orthonormal basis matrices as a tour. To demonstrate this generality, we implement four strategies targeting different analytical tasks, organized into two families.

\paragraph{Hyperdimensional tours.}
These tours explore a single high-dimensional data space and focus on ``what the data looks like from different angles''. We implemented two tours based on spectral decompositions.
The \emph{little tour}~\cite{wickham2011tourr} cycles through projections along successive pairs of components (e.g., PC1-PC2, PC2-PC3, etc.) providing an accessible starting point for any dataset.
The \emph{LE tour} uses Laplacian Eigenmaps~\cite{Belkin2003-ea}, a spectral manifold learning technique, with a cumulative circular basis construction that progressively adds eigenvectors at uniform angular offsets. See Suppl. Sec. 2 for details.

\paragraph{Sequential embedding tours.}
These tours allow for comparison across embedding methods, hyperparameters, and models rather than across directions in a single manifold. They address the question of ``how the picture changes when the lens changes'' and are constructed from sequences of aligned 2D embeddings of the same or one-to-one corresponding data points that serve as keyframes. The embeddings are concatenated so the guided tour interpolates smoothly between them. Here, unlike hyperdimensional tours, only the keyframes are interpretable: the smooth interpolation is an object-constancy aid for tracking points across embeddings, not a claim that intervening frames are meaningful embeddings, which \dtour{} flags with visual cues (Suppl. Fig.~4).
Our \emph{sequential tour} implementation runs a DR method per frame, warm-started from the previous embedding, then Procrustes-aligns and stacks the results (Suppl. Sec. 3). Its special case, the \emph{attraction-repulsion tour}, sweeps Böhm et al.'s~\cite{boehm2022attraction} exaggeration hyperparameter across the spectrum of one neighbor graph.

\subsection{Rendering and Implementation}
\label{sec:implementation}

\dtour{} is available as a TypeScript renderer, React component, and Python-based Anywidget~\cite{anywidget_joss}. Rendering is offloaded to a WebGPU/WebGL2 worker to keep the UI thread free. A separate data worker streams Parquet columns directly to the GPU worker. On an Apple M1 Max MacBook, this architecture sustains smooth playback: $\gtrsim$60 FPS at $\leq$5M, 40 FPS at 10M, and remains usable at 20M points (25 FPS). See Suppl. Sec. 4 for details.

\begin{figure*}
  \centering
  \includegraphics[width=\linewidth]{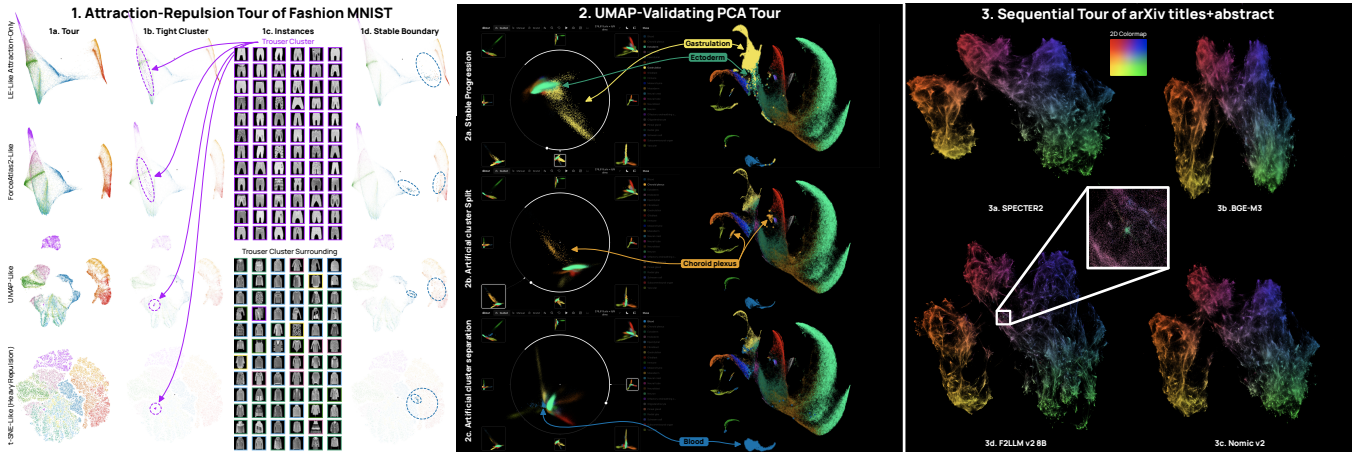}
  \caption{\textbf{Usage Scenarios.} Left: Attraction--repulsion tour of 70K Fashion MNIST images. Middle: UMAP-Validating little PCA tour of 276K single-cell RNA-seq cells. Right: Sequential embedding tour of 3M arXiv titles and abstracts.} \label{fig:usage-scenarios}
\end{figure*}

\section{Usage Scenarios}
\label{sec:usage-scenarios}

We demonstrate \dtour{} in two usage scenarios: (1) gradually revealing structure in high-dimensional data through guided touring and manual manipulation, and (2) validating non-linear DR outputs by touring across embedding methods or models to check whether observed structure is genuine or artifactual. The supplementary video shows the fluid transitions that static figures cannot convey.

\subsection{Gradually Revealing Structure}
\label{sec:gradually-revealing-structure}

No single projection fully captures a high-dimensional manifold. Instead, understanding emerges from viewing multiple projections and the transitions between them.

\paragraph{Attraction-Repulsion Spectrum.}
\label{sec:attraction-repulsion-spectrum}

As shown in \autoref{fig:usage-scenarios}.1, we apply the attraction-repulsion tour to Fashion MNIST~\cite{xiao2017fashion}, sweeping from attraction-only LE (continuous layout) through ForceAtlas2 and UMAP to repulsion-dominated t-SNE (distinct clusters)~\cite{boehm2022attraction}. Scrubbing through the tour (1a) reveals this progression: the continuous LE-like layout gradually separates into the clusters visible in UMAP and sharpened in t-SNE. During guided traversal at the UMAP-like keyframe, we notice a tight cluster (1b) of 96 points embedded among shirts, dresses, and pullovers (1c)---far from the main trouser cluster. Selecting these points and scrubbing back to the ForceAtlas2-like keyframe reveals that they spread across the layout: the tight cluster is an artifact of repulsive forces, not a reflection of genuine data structure. Inspecting the images confirms all 96 are short trousers whose compact silhouette resembles upper-body garments, explaining their misplacement. Yet touring also reveals stability: (1d) boundary points (e.g., boot-like bags bridging the footwear and bag clusters) persist across the entire spectrum, suggesting that boundary placement can be more trustworthy than cluster tightness.

\paragraph{Laplacian Eigenmaps Tour.}
\label{sec:laplacian-eigenmaps-tour}

As shown in \autoref{fig:teaser}, we apply a spectral Fisher variant of the \emph{LE tour}, which orders Laplacian eigenvectors by discriminative power via LDA, to a single-cell dataset of 346K immune cells profiled by CyTOF across 9 surface protein markers from Mair et al.~\cite{mair2022extricating}, with cell-type labels derived from FAUST~\cite{greene2021new}. The tour recovers known immunological hierarchy without specifying which markers matter. The first keyframe (1) separates cells along CD4 versus CD8---the fundamental division between helper and cytotoxic T cells. The second is driven by CD103 and ICOS, distinguishing tissue-resident from activated and regulatory populations---precisely the axis Mair et al. identified as most relevant to tumor-immune differences. Subsequent frames resolve finer structure through markers of T cell regulation (CD25), cytotoxicity (Granzyme), activation (CD38), and exhaustion (Tim3). Notably, CD3---the canonical T cell marker---appears only in late frames with low loading, confirming that the tour correctly assigns minimal weight to constant markers. To go further, we select regulatory T cells (Tregs) during guided tour (2) and switch to manual mode: (3) dragging the ICOS axis separates ICOS-high from ICOS-low Tregs, isolating the tumor-enriched immunosuppressive subset Mair et al. identified as distinguishing cancer from non-malignant inflammation (Suppl. Fig.~1).

\subsection{Validating Embedding Structure}
\label{sec:validating-embedding-structure}

Beyond revealing structure, a second challenge is validating whether patterns in non-linear DR outputs reflect genuine data structure or projection artifacts. \dtour{} addresses this by touring higher-dimensional embeddings or across multiple models.

\paragraph{UMAP-Validating PCA Tour.}
\label{sec:umap-validating-pca-tour}

Single-cell analysis pipelines commonly select highly variable genes, reduce to the top principal components, and embed the resulting PCA space into 2D with UMAP~\cite{becht2019dimensionality}. Because UMAP operates on the PCA output, touring PC pairs provides a natural validation: structure in UMAP but absent from the PCA tour must come from the non-linear embedding. As shown in \autoref{fig:usage-scenarios}.2, we apply a little PCA tour to 276K cells from a developing mouse brain atlas~\cite{lamanno2021molecular}, touring the first 8 principal components alongside a 2D UMAP of the same PCA space. Some structures are stable across both representations: (2a) gastrulation and ectoderm cells form a consistent progression in every PCA keyframe and in UMAP, confirming genuine transcriptional coherence. Other structures diverge: (2b) choroid plexus cells form a single cohesive cluster throughout the PCA tour but split into two distant UMAP clusters, and (2c) blood cells, which UMAP isolates as a disconnected island, show no comparable separation in the PCA tour. These contrasts show that touring PC pairs can distinguish genuine structure from embedding artifacts.

\paragraph{Sentence Embedding Model Comparison.}
\label{sec:sentence-embedding-model-comparison}

To compare embedding models, we construct a sequential tour from 3M arXiv title-abstract embeddings produced by four models spanning 2023--2026: SPECTER2~\cite{singh2023scirepeval}, BGE-M3~\cite{chen2024bgem3}, Nomic Embed Text v2~\cite{nussbaum2025nomic}, and F2LLM-v2-8B~\cite{zhang2026f2llmv2} (top-ranked on clustering benchmarks), each reduced to 2D with default UMAP. Touring across models, which share no common coordinate frame, reveals broad stability (\autoref{fig:usage-scenarios}.3): the overall topical landscape is consistent across all four embeddings despite a 10$\times$ range in model size. However, (3d) F2LLM produces visibly tighter subclusters. A 2D colormap encoding position in the SPECTER2 frame makes structural shifts immediately visible: (3e) one prominent compact cluster in F2LLM disperses across the embedding in all three encoder-based models. Of the ${\sim}$1,200 selected papers, ${\sim}$84\% are physics education research, whose pedagogical language differs sharply from typical arXiv prose. F2LLM appears to cluster these by \emph{discourse style} rather than research topic, while the citation-trained SPECTER2 distributes them among their respective subfields (Suppl. Fig.~2). Thus sequential tours extend embedding validation from inspecting one layout to comparing what models treat as similarity, surfacing behavioral differences no single 2D projection reveals.

\section{Conclusion}
\label{sec:conclusion}

High-dimensional data visualization is hard as no single projection captures a complex manifold's full structure. \dtour{} shows that tours become practical when traversal is fluid and progressive: effortless scrubbing and selection let users gradually build intuition that no static view can provide. We hope \dtour{}'s scalability and availability across Python and web ecosystems spur novel tours and applications that transcend the single 2D embedding.

\acknowledgments{The authors used Anthropic's Claude AI to implement the software and copy-edit the manuscript; the architecture, methodological design, and initial draft are the authors' own. No figures or images were AI-generated.}

\bibliographystyle{abbrv}

\bibliography{template}
\end{document}